\def\mycmd{2}

\if\mycmd1
\documentclass[11pt,onecolumn, draftcls]{IEEEtran}
\else 
\documentclass[lettersize,journal]{IEEEtran}
\fi

\usepackage{ifxetex,ifluatex}
\ifxetex\else\ifluatex\else
\usepackage[latin9]{inputenc}
\fi\fi
\usepackage{tabularx}
\usepackage{threeparttable}

\usepackage{colortbl}    % 표 색상 (선 색상 변경)
\usepackage{pifont}     % 기호 (\ding{})

\usepackage[british]{babel}
\usepackage{multicol}
\usepackage{array,ragged2e}
\usepackage{float}
\usepackage{mathtools}
\usepackage{amsmath}
\usepackage{amsthm}
\usepackage{amssymb}
\usepackage{graphicx}
\usepackage{wasysym}
\usepackage{setspace}
\usepackage{color}
\usepackage{bm}
\usepackage{cases}
\usepackage{makecell}
\usepackage{tikz}
\usepackage{flowchart}
\usepackage{amsfonts}
\usetikzlibrary{matrix,shapes,arrows,positioning,chains}
\usepackage{lmodern,babel,adjustbox,booktabs,multirow}
\usepackage{makecell}
\usepackage{pbox}
\usepackage{epsfig}
\usepackage{subfigure}
\usepackage{tcolorbox}
\usepackage{enumitem}

\definecolor{lightergray}{gray}{0.8}
\definecolor{ForestGreen}{RGB}{34,139,34}

\newcommand{\Xmark}{\textcolor{lightergray}{\ding{55}}}
\newcommand{\mycheck}{\textcolor{ForestGreen}{\ding{51}}}

\allowdisplaybreaks[4]

\makeatletter
\newcommand{\multiline}[1]{%
\begin{tabularx}{\dimexpr\linewidth-\ALG@thistlm}[t]{@{}X@{}}
    #1
\end{tabularx}
}

\floatstyle{ruled}
\newfloat{algorithm}{tbp}{loa}
\providecommand{\algorithmname}{Algorithm}
\floatname{algorithm}{\protect\algorithmname}

\theoremstyle{plain}

\theoremstyle{plain}

\theoremstyle{plain}

\usepackage{epsfig}
\usepackage[caption=false,font=normalsize,labelfont=sf,textfont=sf]{subfig}
\usepackage{cite}
\usepackage{stfloats}
\usepackage{graphicx}
\usepackage{multirow}
\usepackage{array}
\usepackage{graphicx}
\usepackage{epstopdf}                 % auto-convert .eps -> .pdf for pdflatex
\usepackage{times}
\usepackage{algorithm}
\usepackage{algpseudocode}
\theoremstyle{remark}

\newtheorem{remark}{Remark}

\makeatother

\algrenewcommand\algorithmicindent{1.0em}%
\providecommand{\lemmaname}{Lemma}
\providecommand{\propositionname}{Proposition}

\providecommand{\theoremname}{Theorem}
\providecommand{\theoremname}{Definition}
\newcommand{\rom}[1]{\uppercase\expandafter{\romannumeral #1\relax}}

\newcounter{problem}
\newcounter{save@equation}
\newcounter{save@problem}


\numberwithin{save@problem}{subsection}
\numberwithin{save@equation}{subsection}

\makeatletter
\renewcommand\paragraph[1]{%
    \vspace{.1cm}\noindent\textbf{#1}
}
\makeatother

\begin{document}
\title{Dispersion-aware Localization Network for Wideband OFDM Pinching-Antenna Integrated Sensing and Communication Systems}
\author{Hyeonho Noh,~\IEEEmembership{Member,~IEEE} and Hyun Jong Yang,~\IEEEmembership{Senior Member,~IEEE}
\thanks{Hyeonho Noh is with the Department of Information and Communication Engineering, Hanbat National University, Republic of Korea (e-mail: hhnoh@hanbat.ac.kr). Hyun Jong Yang is with the Department of Electrical and Computer Engineering, Seoul National University, Seoul, Korea and also with the Institute of New Media and Communications, Seoul National University, Seoul, Korea (e-mail: hjyang@snu.ac.kr).
}
}

\maketitle
\begin{abstract}\label{abstract}
Pinching-antenna systems (PASS) provide a large effective aperture and substantial path-loss reduction at low hardware cost, making them attractive for integrated sensing and communication (ISAC).  Under wideband OFDM operation, however, the antennas on each waveguide impose nonlinear, position-dependent group delays on the same baseband signal, giving rise to waveguide dispersion that severely degrades range estimation.  To this end, this paper proposes a two-stage ISAC framework comprising communication-aware beamforming and antenna placement followed by a dispersion-aware target localization stage.  A fractional-programming beamformer and an element-wise coordinate-descent placement jointly maximize the downlink sum rate under a sensing beampattern-gain constraint, and the resulting optimized beamformer and placement determine the effective sensing channel used by the localization stage. The proposed \textbf{Dis}\textbf{P}ersion-aware \textbf{L}ocalization \textbf{Net}work (\textbf{DiPL-Net}) detects targets from the received signal via a dispersion-aware score map built on a physics-derived range dictionary. Its convolutional backbone employs dual-kernel residual blocks, each pairing a short kernel matched to the OFDM main lobe with a long kernel matched to the dispersion tail, so as to deconvolve the dispersion and restore a sharp target peak at each true target. Simulation results show substantial localization gains over various sensing baselines while preserving the achievable communication rate.
\end{abstract}


\section{Introduction}
\label{sec:introduction}

The continuing surge in wireless capacity demands has driven a class of flexible antenna architectures whose elements can be actively repositioned to reshape the wireless channel rather than be treated as static~\cite{Liu25_PASS_Tutorial}.  Among them, pinching-antenna systems (PASS) place multiple movable radiating elements---the pinching antennas (PAs)---along each of several dielectric leaky waveguides driven by a single radio-frequency (RF) chain~\cite{Ding25_TCOM_FlexibleAntenna}. Since the waveguides can be routed close to the service area and the PAs can slide along them, PASS realizes a large effective aperture together with substantial path-loss reduction at a fraction of a phased array's hardware cost, while remaining reconfigurable after deployment~\cite{Ding25_TCOM_FlexibleAntenna,Wang25_TCOM}. This same path-loss advantage carries over to the echo link, where it directly boosts the sensing signal-to-noise ratio, making PASS an attractive transmit aperture for integrated sensing and communication (ISAC)~\cite{Hassanien16_TSP_DFRC,Liu22_JSAC_ISAC,Liu20_TCOM_DFRC_Roadmap}.

\begin{figure}[t]
\centering
\includegraphics[width=\columnwidth]{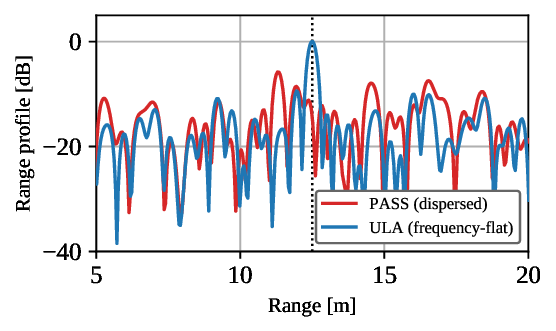}
\caption{Single-target OFDM range profile for a conventional ULA and a pinching-antenna system, with the true target range marked (dotted line).}
\label{fig:intro_range_profile}
\end{figure}

The combination of large aperture, near-target placement, and low hardware cost makes PASS particularly compelling for ISAC~\cite{Liu25_PASS_Tutorial,Ding25_TCOM_FlexibleAntenna}. The most widely studied flexible-antenna architectures for ISAC include reconfigurable intelligent surfaces (RIS), movable antennas (MA), and fluid antennas (FA), but each suffers from limitations that PASS structurally avoids. RIS-aided ISAC incurs a double-fading penalty due to passive reflection \cite{Ndjiongue21_WC}, whereas MA/FA-aided systems remain susceptible to line-of-sight (LoS) blockage because their elements can only be repositioned within a transmitter-confined local region (typically a few wavelengths), which is insufficient to route signals around large-scale obstacles~\cite{Liu25_PASS_Tutorial,Zhu24_CM}. PASS, by contrast, establishes dynamic LoS links and sensing targets through full-length waveguide deployment and freely positioned PAs, making it a compelling transmit architecture for next-generation cellular ISAC where high-rate communication and high-resolution sensing must be delivered concurrently~\cite{Hassanien16_TSP_DFRC,Liu22_JSAC_ISAC,Liu20_TCOM_DFRC_Roadmap}.

\subsection{Related Works}
\label{subsec:related_works}

\paragraph{PASS Hardware and Communication Systems.}
Pinching antennas were introduced as a low-cost, reconfigurable analog of fluid antennas in which dielectric protrusions on a leaky waveguide form radiating elements at adjustable positions~\cite{Liu25_PASS_Tutorial}.  A flexible-antenna perspective on PASS that establishes its single-RF-chain large-aperture advantage was developed in~\cite{Ding25_TCOM_FlexibleAntenna}, downlink rate-maximization for single-waveguide PASS was studied in~\cite{Xu25_WCL_PASS_RateMax}, and PASS has subsequently been applied to physical-layer security~\cite{Wang26_WCL_PASS_PLS} and covert communications~\cite{Jiang26_JSAC_PASS_Covert}. Collectively, these studies indicate PASS as a viable transmit aperture for next-generation cellular systems.

\paragraph{PASS for ISAC.}
A rapidly growing body of work investigates PASS as a transmit aperture for ISAC.  Wireless sensing via PASS was characterized as a sensing-only baseline in~\cite{Wang25_WCL_PASSsensing}.  Joint PA placement and transmit-power optimization for PASS-aided ISAC was studied in~\cite{Qin25_WCL_PASS_ISAC} via reinforcement learning under communication-rate and energy constraints.  Fundamental information-theoretic limits of PASS-ISAC under sensing-centric, communication-centric, and Pareto-optimal beamforming objectives were analyzed in~\cite{Wang25_CL_PASS_ISAC}, and a multi-waveguide PASS-ISAC framework with jointly optimized transmit and receive PAs under rate, signal-to-noise ratio (SNR), power, and placement constraints was proposed in~\cite{Mao26_TWC_PASS_ISAC} using fine-tuning approximations and successive convex approximation.  Most directly related to this work, Li~\textit{et al.}~\cite{Li26_TWC_PASS_ISAC} proposed a CRB-based PASS-ISAC framework that optimizes PA positions to minimize the Cram\'er--Rao bound (CRB) of the target parameters under a communication sum-rate constraint. Near-field PASS-ISAC extensions to joint downlink/uplink sensing have also been explored~\cite{Li24_JSAC_NearFieldISAC}, and broader ISAC waveform-design road maps are surveyed in~\cite{Wei23_IoT_ISAC_Survey,Liu20_TCOM_DFRC_Roadmap,Hassanien16_TSP_DFRC,Liu22_JSAC_ISAC}.  These works collectively establish the value of jointly leveraging PA placement and beamforming for ISAC but are confined to the narrowband regime. The wideband OFDM-PASS regime, in which the in-waveguide propagation phase varies nonlinearly with frequency, has been examined only very recently for frequency-selective transmit beamforming~\cite{Xiao25_OFDM_PASS} and channel estimation~\cite{Chen25_OFDM_PASS_CE}, leaving the corresponding dispersion-aware target localization problem unresolved.  Table~\ref{tab:compare_pass_isac} summarizes how the proposed scheme differs from these prior PASS-ISAC works along the bandwidth, waveguide-architecture, detector-type, and metric axes most relevant to this work.

\begin{table}[t]
\centering
\caption{Comparison of the proposed scheme to prior PASS-ISAC works.}
\label{tab:compare_pass_isac}
\adjustbox{width=\columnwidth}{
\begin{tabular}{ccccc}
\toprule
\textbf{Ref.} & \textbf{Bandwidth} & \textbf{Waveguides} & \textbf{PASS dispersion} & \textbf{Detector type} \\
\midrule[\heavyrulewidth]
\arrayrulecolor{lightergray}
\cite{Wang25_WCL_PASSsensing} & Narrowband & Multiple & \Xmark & Model-based \\ \cline{1-5}
\cite{Qin25_WCL_PASS_ISAC} & Narrowband & Single & \Xmark & Model-based \\ \cline{1-5}
\cite{Wang25_CL_PASS_ISAC} & Narrowband & Single & \Xmark & Model-based \\ \cline{1-5}
\cite{Li24_JSAC_NearFieldISAC} & Narrowband & Single & \Xmark & Model-based \\ \cline{1-5}
\cite{Mao26_TWC_PASS_ISAC} & Narrowband & Multiple & \Xmark & Model-based \\ \cline{1-5}
\cite{Li26_TWC_PASS_ISAC} & Narrowband & Multiple & \Xmark & Model-based \\
\arrayrulecolor{black}
\midrule[\heavyrulewidth]
\textbf{Ours} & \textbf{Wideband OFDM} & \textbf{Multiple} & \mycheck & \textbf{Model + Learning} \\
\bottomrule
\end{tabular}}
\end{table}

\subsection{Challenges}
\label{subsec:challenges}

Current 5G New Radio and 6G ISAC standardization activities have largely converged on orthogonal frequency-division multiplexing (OFDM) as the waveform of choice, since range resolution scales inversely with the signal bandwidth and OFDM blends naturally with the cellular downlink~\cite{Pucci22_JSAC_NR_ISAC, Noh23_TVT}.  Combining OFDM with PASS, however, exposes a wideband sensing impairment absent from the narrowband regime studied by existing PASS-ISAC works~\cite{Li26_TWC_PASS_ISAC,Liu22_TSP_CRBISAC,Liu20_TSP_JointBF}.  Across the wide OFDM band, the in-waveguide propagation phase varies nonlinearly with frequency, so the multiple antennas sharing each waveguide combine with a different phase relationship on every subcarrier and the effective transmit channel changes from one subcarrier to the next.  Since OFDM range estimation extracts the target delay from a linear-in-frequency phase progression~\cite{Sturm11_Proc}, this dispersion corrupts that linearity. As illustrated in Fig.~\ref{fig:intro_range_profile}, the range echo from a single target is smeared into a long-tailed point spread function (PSF) instead of the clean sinc obtained on a uniform linear array (ULA), and the localization error inflates far beyond the OFDM range resolution.
Motivated by these unresolved challenges, this study poses a key research question:
\begin{tcolorbox}[colframe=black, colback=white, height=1.1cm, boxrule=0.4mm]
\begin{center}
\vspace{-0.137cm}
\textit{\textbf{How can a detector mitigate dispersion while maintaining low computation cost?}}
\end{center}
\end{tcolorbox}
\noindent A purely data-driven network trained on the raw echo requires extensive training data and generalizes poorly, as the dispersion spreads each target across range and entangles it with clutter~\cite{He19_WC}.  To address this, the proposed DiPL-Net adopts a physics-guided design that embeds the dispersion model into the detector input and architecture.

\subsection{Contributions}
\label{subsec:contributions}

This paper develops a two-stage wideband OFDM-PASS ISAC framework. Phase~1 jointly optimizes the beamformers and PA positions for downlink communication subject to a sensing beampattern constraint. Phase~2 performs dispersion-aware target localization using the optimized sensing channel. The Phase-2 detector is termed the \textbf{Di}s\textbf{P}ersion-aware \textbf{L}ocalization \textbf{Net}work (\textbf{DiPL-Net}). It combines a model-based input representation with a physics-guided convolutional architecture. The main contributions are summarized as follows.
\begin{enumerate}[leftmargin=*]
    \item \textbf{Wideband OFDM-PASS sensing model.} Existing PASS-ISAC formulations mainly consider narrowband channels~\cite{Li26_TWC_PASS_ISAC,Liu22_TSP_CRBISAC}, whereas this paper formulates a wideband OFDM-PASS sensing model in which the in-waveguide propagation constant induces a frequency-dependent effective transmit channel. Closed-form expressions are derived for the dispersion-aware effective channel and the subcarrier beampattern gain, which together characterize the range-domain distortion caused by waveguide dispersion.

    \item \textbf{Communication-oriented beamforming and PA placement with a sensing-performance guarantee.} A downlink sum-rate maximization problem is formulated under transmit-power, target beampattern-gain, and PA feasibility constraints. The proposed solver alternates between two updates: a closed-form fractional-programming beamformer update with Sherman--Morrison-updated multipliers~\cite{Shen18_FP}, and an element-wise coordinate-descent update for the PA positions. The optimized PA positions and transmit beamformer define the effective sensing channel used by DiPL-Net.

    \item \textbf{Physics-guided localization network.} DiPL-Net incorporates the dispersion model through its input representation. The received signal is first projected onto a score map constructed from a dispersion-aware range dictionary and a spatial-spectrum angle prior. This model-based front end encodes the deterministic wideband response before learning is applied. The convolutional backbone uses dual range-axis kernels: a short kernel matched to the OFDM main lobe and a long kernel matched to the dispersion-tail support. A dual-head decoder then separates grid-bin detection from sub-bin offset regression.

    \item \textbf{Integrated two-stage ISAC validation.} Simulations under a millimeter-wave wideband ISAC setting show that the proposed PA position and beamforming optimization improves the downlink sum rate while satisfying the sensing beampattern constraint. The results also verify that DiPL-Net reduces localization error relative to multiple sensing-detector baselines. A factorial ablation study quantifies the contribution of each component.
\end{enumerate}

\paragraph{Notation.} Scalars, vectors, and matrices are denoted by italic, bold lower-case, and bold upper-case letters, respectively. The operators $(\cdot)^*$, $(\cdot)^\mathrm{T}$, and $(\cdot)^\mathrm{H}$ denote complex conjugation, transpose, and conjugate transpose. For a matrix $\mathbf{A}$, $\mathbf{A}^{-1}$, $|\mathbf{A}|$, and $[\mathbf{A}]_{i,j}$ denote its inverse, determinant, and $(i,j)$-th entry, while $[\mathbf{a}]_i$ denotes the $i$-th entry of a vector $\mathbf{a}$. The symbols $\mathbf{I}_N$ and $\mathbf{0}$ denote the $N\times N$ identity matrix and an all-zero vector or matrix of proper dimension. $\mathbb{R}^{m\times n}$ and $\mathbb{C}^{m\times n}$ denote the real and complex matrix spaces, and $\mathcal{CN}(\boldsymbol{\mu},\boldsymbol{\Sigma})$ denotes a circularly symmetric complex Gaussian distribution with mean $\boldsymbol{\mu}$ and covariance $\boldsymbol{\Sigma}$. Unless otherwise specified, $\|\cdot\|$ denotes the Euclidean norm, $\|\cdot\|_1$ denotes the $\ell_1$ norm, and $\mathrm{Re}[\cdot]$ denotes the real part.

\section{System Model}
\label{sec:system_model}

\subsection{System Geometry and Notation}
\label{subsec:sm_geometry}

\begin{figure}[t]
\centering
\includegraphics[width=\columnwidth]{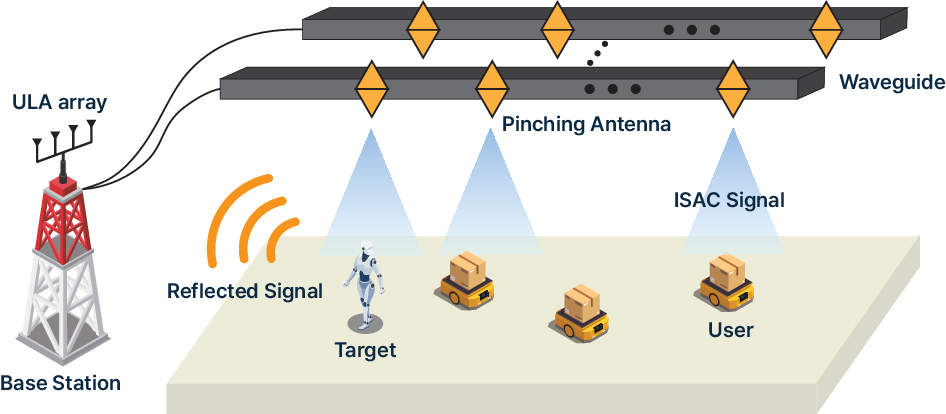}
\caption{Proposed system model.}
\label{fig:system_model}
\end{figure}

A PASS-aided ISAC base station (BS) serves $K$ single-antenna users while sensing $J$ targets using the separated Tx/Rx architecture in Fig.~\ref{fig:system_model}. The Tx side comprises $N$ dielectric waveguides, each carrying $M$ PAs, and the co-located Rx side uses an $N_\mathrm{R}$-element ULA as in the self-interference-avoiding layout of~\cite{Li26_TWC_PASS_ISAC}. Waveguide $n$ is placed at height $d_h$ and extends along the $x$-axis from the feed point $\boldsymbol{\psi}_0^{(n)} \!=\! [0, y^{(n)}, d_h]^\mathrm{T}$. Its $m$-th PA is located at $\boldsymbol{\psi}_m^{(n)} \!=\! [x_m^{(n)}, y^{(n)}, d_h]^\mathrm{T}$, where
\begin{align}
    0 &\le x_m^{(n)} \le L, \label{eq:PA_box}\\
    x_m^{(n)} - x_{m-1}^{(n)} &\ge \delta, \quad \forall\, m \in \mathcal{M},\; n \in \mathcal{N}, \label{eq:PA_sep}
\end{align}
We denote the waveguide length by $L$ and set the minimum inter-PA spacing to $\delta = \lambda_\mathrm{c}/2$ for wavelength $\lambda_\mathrm{c}$. The PAs are ordered along each waveguide as $x_1^{(n)} < x_2^{(n)} < \cdots < x_M^{(n)}$, and their positions are collected in $\mathbf{X} \!=\! \{x_m^{(n)}\}_{m \in \mathcal{M}, n \in \mathcal{N}}$. User and target locations are denoted by $\mathbf{u}_k \!\in\! \mathbb{R}^3$ and $\mathbf{p}_j \!\in\! \mathbb{R}^3$, respectively. Each sensing snapshot is formed from one downlink OFDM symbol spanning $N_\mathrm{c}$ subcarriers around carrier frequency $f_\mathrm{c}$. With subcarrier spacing $\Delta f$, the occupied bandwidth is $B = N_\mathrm{c}\Delta f$, and $f_i$ denotes the frequency of subcarrier $i$. The following indices are used throughout the paper:
\begin{itemize}[leftmargin=*,itemsep=0pt,topsep=2pt]
\item $n \in \mathcal{N}\!=\!\{1,\ldots,N\}$ -- waveguide;
\item $m \in \mathcal{M}\!=\!\{1,\ldots,M\}$ -- PA on a single waveguide;
\item $i \in \mathcal{I}\!=\!\{0,\ldots,N_\mathrm{c}\!-\!1\}$ -- OFDM subcarrier;
\item $k \in \mathcal{K}\!=\!\{1,\ldots,K\}$ -- communication user;
\item $j \in \mathcal{J}\!=\!\{1,\ldots,J\}$ -- sensing target;
\item $r \in \{1,\ldots,N_\mathrm{R}\}$ -- BS Rx-ULA antenna.
\end{itemize}

\subsection{PASS Channel and Tx Signal Model}
\label{subsec:pass_channel}

The Tx-side channel from the waveguide-$n$ feed to a point $\mathbf{p}$ via PA $m$ is modeled as the cascade of guided propagation inside the waveguide and free-space propagation after radiation. The guided path has length $x_m^{(n)}$, attenuation $\alpha_g(f)$, and propagation constant $\beta_g(f) \!=\! (2\pi/c)\sqrt{f^2 - f_\mathrm{cut}^2}$ for frequencies above the cutoff frequency $f_\mathrm{cut}$. The free-space hop has length $r_m^{(n)}(\mathbf{p}) \!=\! \|\boldsymbol{\psi}_m^{(n)} \!-\! \mathbf{p}\|$ and aperture gain $G_\mathrm{a}(f) \!=\! c^2/(16\pi^2 f^2)$. The resulting channel is expressed as
\begin{align}\label{eq:per_PA_channel}
    h_m^{(n)}(\mathbf{p}, f) &= \overbrace{\sqrt{\rho}\, e^{-\alpha_g(f)\, x_m^{(n)} - j\beta_g(f)\, x_m^{(n)}}}^{\text{in-waveguide}} \nonumber \\ 
    &\hspace{50pt} \cdot \underbrace{\frac{\sqrt{G_\mathrm{a}(f)}}{r_m^{(n)}(\mathbf{p})}\, e^{-j 2\pi f\, r_m^{(n)}(\mathbf{p})/c}}_{\text{free space}},
\end{align}
where $\rho = 1/M$ is the equal-power-split coefficient \cite{Li26_TWC_PASS_ISAC}.  The nonlinear dependence of $\beta_g$ on $f$ is the source of waveguide dispersion: the phase shifts $\beta_g(f_i)\,x_m^{(n)}$ vary nonlinearly across subcarriers and accumulate over many full rotations across the band at typical millimeter-wave PA positions.

Since all $M$ PAs on the same waveguide share a single RF chain and radiate the same baseband signal, the BS-Tx-side effective PASS channel to $\mathbf{p}$ on subcarrier $i$ is given by
\begin{align}\label{eq:effective_channel}
    \big[\tilde{\mathbf{h}}(\mathbf{X}, f_i, \mathbf{p})\big]_n = \sum_{m=1}^{M} h_m^{(n)}(\mathbf{p}, f_i), \quad n \in \mathcal{N}.
\end{align}
On subcarrier $i$, the BS maps the $K$ user streams to the $N$ waveguide feeds through waveguide-domain beamforming. For the unit-variance data symbol $s_k^{(i)}$ and beamformer $\mathbf{w}_k^{(i)} \in \mathbb{C}^N$ of user $k$, let $\mathbf{W}^{(i)} \!=\! [\mathbf{w}_1^{(i)}, \ldots, \mathbf{w}_K^{(i)}]$ and $\mathbf{s}^{(i)} \!=\! [s_1^{(i)}, \ldots, s_K^{(i)}]^\mathrm{T}$. The corresponding feed-input vector is represented as
\begin{align}\label{eq:tx_vec}
    \mathbf{x}^{(i)} = \mathbf{W}^{(i)}\, \mathbf{s}^{(i)} \in \mathbb{C}^N.
\end{align}

\subsection{Downlink Communication Model}
\label{subsec:communication_model}

Let $\tilde{\mathbf{h}}_k^{(i)} = \tilde{\mathbf{h}}(\mathbf{X}, f_i, \mathbf{u}_k) \in \mathbb{C}^N$ denote the effective channel to user $k$ on subcarrier $i$, obtained by evaluating \eqref{eq:effective_channel} at $\mathbf{u}_k$. The received sample at user $k$ on subcarrier $i$ is given by
\begin{align}\label{eq:user_received}
y_k^{(i)} = \underbrace{(\tilde{\mathbf{h}}_k^{(i)})^\mathrm{H} \mathbf{w}_k^{(i)}\, s_k^{(i)}}_{\text{desired signal}} + \underbrace{\sum_{\ell\ne k}\!(\tilde{\mathbf{h}}_k^{(i)})^\mathrm{H} \mathbf{w}_\ell^{(i)}\, s_\ell^{(i)}}_{\text{multi-user interference}} + z_k^{(i)},
\end{align}
with $z_k^{(i)} \!\sim\! \mathcal{CN}(0,\sigma_0^2)$. Then, the signal-to-interference-plus-noise ratio (SINR) is written as
\begin{align}\label{eq:sinr}
    \gamma_k^{(i)} = \frac{\big|(\tilde{\mathbf{h}}_k^{(i)})^\mathrm{H} \mathbf{w}_k^{(i)}\big|^2}{\sum_{\ell \in \mathcal{K} \backslash \{k\}} \big|(\tilde{\mathbf{h}}_k^{(i)})^\mathrm{H} \mathbf{w}_\ell^{(i)}\big|^2 + \sigma_0^2},
\end{align}
and the achievable downlink sum rate is $R_\mathrm{sum} = \Delta f \sum_{i=0}^{N_\mathrm{c}-1} \sum_{k\in\mathcal{K}} \log_2(1+\gamma_k^{(i)})$.

\subsection{Sensing Model}
\label{subsec:sensing_model}

\paragraph{Channel model.}  The PASS-to-target and PASS-to-user effective channels follow a Rician model. The deterministic LoS component is given by \eqref{eq:effective_channel}, while the diffuse component is modeled as an NLoS perturbation:
\begin{align}
\tilde{\mathbf{h}} = \sqrt{\frac{K_\mathrm{R}}{K_\mathrm{R}+1}}\,\tilde{\mathbf{h}}_\mathrm{LoS} + \sqrt{\frac{1}{K_\mathrm{R}+1}}\,\tilde{\mathbf{h}}_\mathrm{NLoS},
\end{align}
where $K_\mathrm{R}$ is the Rician factor and $\tilde{\mathbf{h}}_\mathrm{NLoS} \sim \mathcal{CN}(\mathbf{0}, \sigma_\mathrm{h}^2 \mathbf{I}_N)$ with mean NLoS power $\sigma_\mathrm{h}^2$.
% Since the diffuse component has zero mean and is independent of the PA positions $\mathbf{X}$, the Phase-1 optimization gradient with respect to $\mathbf{X}$ is well-defined and the optimal beamformer at the mean channel is the LoS-only solution scaled by $\sqrt{K_\mathrm{R}/(K_\mathrm{R}+1)}$.

\paragraph{Received sensing signal.} Because the BS Tx and Rx arrays are co-located, the Tx-to-target and target-to-Rx propagation distances are both denoted by $R_j = \|\mathbf{p}_j - \mathbf{p}_\mathrm{BS}\|$. Assuming that the cyclic prefix absorbs the maximum round-trip delay, the reflection of the user-serving OFDM signal in \eqref{eq:tx_vec} by the $J$ targets yields the Rx-ULA observation matrix $\mathbf{Y} \in \mathbb{C}^{N_\mathrm{R} \times N_\mathrm{c}}$, whose entries are written as
\begin{align}\label{eq:y_matrix}
[\mathbf{Y}]_{r, i} &= \sum_{j=1}^{J} \bar\alpha_j\, [\mathbf{a}_\mathrm{R}(\theta_j)]_r\, e^{-j 2\pi i\Delta f\cdot 2R_j/c}\, \big(\tilde{\mathbf{h}}_j^{(i)}\big)^\mathrm{H}\, \mathbf{W}^{(i)}\mathbf{s}^{(i)} \nonumber \\ & \hspace{160pt}+ [\mathbf{Z}]_{r, i}.
\end{align}
In \eqref{eq:y_matrix}, $\tilde{\mathbf{h}}_j^{(i)} = \tilde{\mathbf{h}}(\mathbf{X}, f_i, \mathbf{p}_j)$ denotes the Tx-side effective channel to target $j$, and the Rx-ULA steering vector at AoA $\theta_j$ is represented as
\begin{align}
\mathbf{a}_\mathrm{R}(\theta_j) =
\big[\,1,\, e^{-j 2\pi \Delta\sin\theta_j},\, \ldots,\, e^{-j 2\pi (N_\mathrm{R}-1)\Delta\sin\theta_j}\,\big]^\mathrm{T},
\end{align}
where $\Delta$ is the antenna spacing normalized by the wavelength, $e^{-j 2\pi i\Delta f\cdot 2R_j/c}$ captures the round-trip delay phase, $\bar\alpha_j = e^{j\phi_j}\sqrt{\sigma_j}/R_j\cdot e^{-j 2\pi f_\mathrm{c}\,2R_j/c}$ absorbs the RCS $\sigma_j$ and carrier-frequency round-trip phase, and $[\mathbf{Z}]_{r, i} \sim \mathcal{CN}(0, \sigma_s^2)$ is the receiver noise.

Correlating the observation $[\mathbf{Y}]_{:, i}$ with the target-$j$ steering vector $\mathbf{a}_\mathrm{R}(\theta_j)$ gives the received subcarrier-domain signal
\begin{align}\label{eq:y_after_rxBF}
y_j^{(i)} = \tilde\alpha_j\, e^{-j 2\pi i\Delta f\cdot 2R_j/c}\, \big(\tilde{\mathbf{h}}_j^{(i)}\big)^\mathrm{H}\, \mathbf{W}^{(i)}\,\mathbf{s}^{(i)} + z_j^{(i)},
\end{align}
where $\tilde\alpha_j = \frac{1}{\sqrt{N_\mathrm{R}}}\,\bar\alpha_j\,\mathbf{a}_\mathrm{R}^\mathrm{H}(\theta_j)\,\mathbf{a}_\mathrm{R}(\theta_j) = \sqrt{N_\mathrm{R}}\,\bar\alpha_j$ is the effective reflectivity at the target-$j$ angle, and $z_j^{(i)}$ collects the receiver noise together with the residual crosstalk from the other $J-1$ targets, governed by the steering correlations $\mathbf{a}_\mathrm{R}^\mathrm{H}(\theta_j)\,\mathbf{a}_\mathrm{R}(\theta_{j'})$, $j'\ne j$. Since both $\mathbf{W}^{(i)}$ and $\mathbf{s}^{(i)}$ are fully known at the BS, the received signal in \eqref{eq:y_after_rxBF} can be written as
\begin{align}\label{eq:symbol_divided_echo}
y_j[i] = \tilde\alpha_j\, H_\mathrm{ch}(f_i;\mathbf{p}_j)\, e^{-j 2\pi i\Delta f\cdot 2R_j/c} + \tilde z_j[i],
\end{align}
with the data-aware effective Tx-side sensing channel
\begin{align}\label{eq:Hch_def}
H_\mathrm{ch}(f_i;\mathbf{p}) = \big(\tilde{\mathbf{h}}(\mathbf{X}, f_i, \mathbf{p})\big)^\mathrm{H} \mathbf{x}^{(i)}
\end{align}
absorbing all PASS-induced wideband effects together with the known per-user data symbols, and $\tilde z_j[i]\!=\! z_j^{(i)}$.

\paragraph{Sensing metric.}  Under independent and identically distributed (i.i.d.) unit-variance data symbols, the expected Tx-side beampattern (BP) gain delivered at target $j$ on subcarrier $i$ is defined as
\begin{align}\label{eq:illumination}
    |G_j^{(i)}|^2 = \sum_{k\in\mathcal{K}} \big|(\tilde{\mathbf{h}}_j^{(i)})^\mathrm{H}\, \mathbf{w}_k^{(i)}\big|^2.
\end{align}

\begin{remark}
An inverse FFT of the symbol-divided echo $y_j[i]$ in \eqref{eq:symbol_divided_echo} localizes the target only if the channel term $H_\mathrm{ch}(f_i;\mathbf{p}_j)$ has linear phase in the subcarrier index $i$. This condition holds for a conventional ULA, for which the FFT recovers a sharp peak at $R_j$. In PASS, the nonlinear-in-frequency propagation constant $\beta_g(f)$ bends the phase of $H_\mathrm{ch}(f_i;\mathbf{p}_j)$ across subcarriers. The resulting long-tailed PSF can obscure the true range peak.
\end{remark}

\section{Phase 1: Joint PA Placement and Beamforming Optimization}
\label{sec:phase1}

Phase~1 jointly optimizes the PA positions $\mathbf{X}$ and the transmit beamformers $\{\mathbf{W}^{(i)}\}$ to maximize the downlink sum rate under total-power, target beampattern-gain, and PA feasibility constraints. The use of a communication-rate objective together with explicit sensing-side beampattern constraints is aligned with the broader joint communication/radar beamforming literature~\cite{DCFNet_TWC2026, Liu20_TSP_JointBF,Liu22_TSP_CRBISAC}, while the present problem additionally accounts for PASS placement feasibility and wideband OFDM dispersion. The proposed alternating optimization (AO) combines a closed-form fractional-programming (FP) beamformer update for each subcarrier with an element-wise coordinate-descent PA-position update over a discrete grid.

\subsection{Problem Formulation}
\label{subsubsec:p1_problem}

To support target detectability, the Tx-side beampattern gain toward each candidate target location $\mathbf{p}_j$ is constrained to exceed $P_\mathrm{req}$~\cite{Li26_TWC_PASS_ISAC,Noh25_iotj}. The Phase-1 optimization problem is then formulated as
\begin{subequations}\label{eq:p1_problem}
\begin{align}
    \max_{\mathbf{X},\,\{\mathbf{w}_k^{(i)}\}}\; & R_\mathrm{sum}(\mathbf{X},\mathbf{W}) \label{eq:p1_obj}\\
    \mathrm{s.t.}\;\;& \sum_{i=0}^{N_\mathrm{c}-1}\sum_{k\in\mathcal{K}}\|\mathbf{w}_k^{(i)}\|^2 \le P_\mathrm{tot}, & \label{eq:p1_power}\\
    & |G_j^{(i)}|^2 \ge P_\mathrm{req}, & \forall i,\, \forall j \in \mathcal{J}, \label{eq:p1_sense}\\
    & \mathbf{X} \in \mathcal{X}, \label{eq:p1_pa}
\end{align}
\end{subequations}
where $\mathcal{X} \!=\! \{\mathbf{X} : 0 \!\le\! x_m^{(n)} \!\le\! L,\, x_{m+1}^{(n)} - x_m^{(n)} \!\ge\! \delta\}$ is the PA feasible set. Constraint \eqref{eq:p1_power} limits the total transmit power across all subcarriers, constraint \eqref{eq:p1_sense} imposes the sensing beampattern floor, and constraint \eqref{eq:p1_pa} confines each PA to its waveguide while enforcing the minimum spacing $\delta$.

Problem~\eqref{eq:p1_problem} is non-convex because the PA positions and transmit beamformers are coupled through the dispersion-aware effective channel $\tilde{\mathbf{h}}(\mathbf{X}, f_i, \cdot)$, while the PA positions enter the channel nonlinearly through the in-waveguide propagation phase. The sensing constraint \eqref{eq:p1_sense} also lower-bounds the quadratic gain $|G_j^{(i)}|^2$, which yields a non-convex feasible region. These properties motivate the proposed alternating optimization.
\subsection{Transmit Beamformer Optimization}
\label{subsubsec:w_subproblem}
With the PA positions fixed, the optimization problem in \eqref{eq:p1_problem} is first solved to obtain the transmit beamformer $\{\mathbf{w}_k^{(i)}\}$.

\paragraph{Step 1: FP transforms on the rate.}  Following \cite{Shen18_FP,DCFNet_TWC2026}, each $\log_2(1+\gamma_k^{(i)})$ admits the Lagrangian dual representation
\begin{align}\label{eq:LDT_p1}
    \log_2(1+\gamma_k^{(i)}) = \!\!\max_{\eta_k^{(i)}\ge 0}\!\big[\log_2(1+\eta_k^{(i)}) - \eta_k^{(i)} + \tfrac{(1+\eta_k^{(i)})\gamma_k^{(i)}}{1+\gamma_k^{(i)}}\big],
\end{align}
attained at $\hat\eta_k^{(i)} \!=\! \gamma_k^{(i)}$; holding $\{\eta_k^{(i)}\}$ fixed, the quadratic transform on $\gamma_k^{(i)}$ gives
\begin{align}\label{eq:QT_p1}
\frac{(1+\eta_k^{(i)})\gamma_k^{(i)}}{1+\gamma_k^{(i)}} = \max_{\xi_k^{(i)}} & \;2\sqrt{1+\eta_k^{(i)}}\,\mathrm{Re}\big[\xi_k^{(i)*}\!(\tilde{\mathbf{h}}_k^{(i)})^\mathrm{H}\mathbf{w}_k^{(i)}\big]\nonumber\\
&\hspace{-20pt} -|\xi_k^{(i)}|^2\!\Big(\!\sum_{p\in\mathcal{K}}|(\tilde{\mathbf{h}}_k^{(i)})^\mathrm{H}\mathbf{w}_p^{(i)}|^2 + \sigma_0^2\Big),
\end{align}
attained at
\begin{align}\label{eq:xi_opt_p1}
\hat\xi_k^{(i)} = \frac{\sqrt{1+\eta_k^{(i)}}\,(\tilde{\mathbf{h}}_k^{(i)})^\mathrm{H}\mathbf{w}_k^{(i)}}{\sum_{p\in\mathcal{K}}|(\tilde{\mathbf{h}}_k^{(i)})^\mathrm{H}\mathbf{w}_p^{(i)}|^2 + \sigma_0^2}.
\end{align}
With ${\eta_k^{(i)}, \xi_k^{(i)}}$ fixed, the objective becomes quadratic in ${\mathbf{w}_k^{(i)}}$, yielding a convex optimization problem.

\paragraph{Step 2: Closed-form beamformer via KKT.}  Let $\tilde{R}_\mathrm{sum}(\mathbf{W})$ denote this FP surrogate of $R_\mathrm{sum}$, that is, the sum of the quadratic transform \eqref{eq:QT_p1} over all users and subcarriers.  Attaching a single multiplier $\lambda\!\ge\!0$ to the total-power constraint \eqref{eq:p1_power} and target multipliers $\{\mu_j^{(i)}\!\ge\!0\}_{j\in\mathcal{J}}$ to the beampattern-gain constraints \eqref{eq:p1_sense} gives the Lagrangian
\begin{align}\label{eq:lagrangian_p1}
\mathcal{L} = \tilde{R}_\mathrm{sum}(\mathbf{W}) &- \lambda\Big(\sum_{i=0}^{N_\mathrm{c}-1}\sum_{k\in\mathcal{K}}\|\mathbf{w}_k^{(i)}\|^2 - P_\mathrm{tot}\Big) \nonumber \\
&+ \sum_{i=0}^{N_\mathrm{c}-1}\sum_{j\in\mathcal{J}}\mu_j^{(i)}\big(|G_j^{(i)}|^2 - P_\mathrm{req}\big).
\end{align}
Setting the Wirtinger gradient~\cite{Brandwood83_Wirtinger} $\partial\mathcal{L}/\partial\mathbf{w}_k^{(i)*} \!=\! \mathbf{0}$ yields the closed form
\begin{align}\label{eq:wk_closed_form_p1}
\mathbf{w}_k^{(i)} = \big(\mathbf{A}^{(i)}\big)^{-1}\mathbf{b}_k^{(i)},
\end{align}
with the regularizing matrix
\begin{align}\label{eq:A_p1}
\mathbf{A}^{(i)} &\!=\! \sum_{p\in\mathcal{K}}|\xi_p^{(i)}|^2 \tilde{\mathbf{h}}_p^{(i)}(\tilde{\mathbf{h}}_p^{(i)})^\mathrm{H} \!+\! \lambda\mathbf{I}_N - \sum_{j\in\mathcal{J}} \mu_j^{(i)}\, \tilde{\mathbf{h}}_j^{(i)}(\tilde{\mathbf{h}}_j^{(i)})^\mathrm{H}
\end{align}
and $\mathbf{b}_k^{(i)} = \xi_k^{(i)}\sqrt{1+\eta_k^{(i)}}\,\tilde{\mathbf{h}}_k^{(i)}$.
Note that the Lagrange multiplier $\lambda$ is chosen large enough that every $\mathbf{A}^{(i)} \succ \mathbf{0}$; in practice, the bisection in Step~3 starts at a $\lambda$ value that dominates the negative semidefinite contribution of the target rank-one terms across all subcarriers, so each $\mathbf{A}^{(i)}$ remains positive definite throughout the optimization and $(\mathbf{A}^{(i)})^{-1}$ is well defined.

\paragraph{Step 3: Solution for $\lambda$.} Let $\mathbf{A}_0^{(i)}$ denote the $\lambda$-independent part of \eqref{eq:A_p1},
\begin{align}\label{eq:M_p1}
\mathbf{A}_0^{(i)} = \sum_{p\in\mathcal{K}}|\xi_p^{(i)}|^2 \tilde{\mathbf{h}}_p^{(i)}(\tilde{\mathbf{h}}_p^{(i)})^\mathrm{H} - \sum_{j\in\mathcal{J}} \mu_j^{(i)}\, \tilde{\mathbf{h}}_j^{(i)}(\tilde{\mathbf{h}}_j^{(i)})^\mathrm{H},
\end{align}
so that $\mathbf{A}^{(i)} = \mathbf{A}_0^{(i)} + \lambda\mathbf{I}_N$.  If every $\mathbf{A}_0^{(i)} \succ \mathbf{0}$ and the induced power $\sum_{i=0}^{N_\mathrm{c}-1}\sum_{k\in\mathcal{K}}\|\mathbf{w}_k^{(i)}\|^2$ evaluated at $\lambda = 0$ does not exceed $P_\mathrm{tot}$, it holds $\lambda^\star = 0$. Otherwise the power constraint holds with equality.  Let $\mathbf{A}_0^{(i)} = \mathbf{V}^{(i)}\boldsymbol{\Lambda}^{(i)}(\mathbf{V}^{(i)})^\mathrm{H}$ be the eigendecomposition; then $\mathbf{w}_k^{(i)} = \mathbf{V}^{(i)}(\boldsymbol{\Lambda}^{(i)}+\lambda\mathbf{I}_N)^{-1}(\mathbf{V}^{(i)})^\mathrm{H}\mathbf{b}_k^{(i)}$, and the total-power equality reduces to
\begin{align}\label{eq:lambda_search}
\sum_{i=0}^{N_\mathrm{c}-1}\sum_{n=1}^{N} \frac{[\boldsymbol{\Upsilon}^{(i)}]_{nn}}{\big([\boldsymbol{\Lambda}^{(i)}]_{nn}+\lambda\big)^2} = P_\mathrm{tot},
\end{align}
with $\boldsymbol{\Upsilon}^{(i)} = (\mathbf{V}^{(i)})^\mathrm{H}\big(\sum_{k\in\mathcal{K}}\mathbf{b}_k^{(i)}(\mathbf{b}_k^{(i)})^\mathrm{H}\big)\mathbf{V}^{(i)}$.  The left-hand side is strictly decreasing in $\lambda$ for $\lambda > -\min_{i,n}[\boldsymbol{\Lambda}^{(i)}]_{nn}$, so the unique $\lambda^\star$ can be obtained via a one-dimensional search.

\paragraph{Step 4: Solution for $\{\mu_j^{(i)}\}$.} With the fixed $\lambda$ and the remaining multipliers, $\mu_j^{(i)}$ is updated cyclically over $j\in\mathcal{J}$ until convergence.  Define $\mathbf{E}_j^{(i)}$ as
\begin{align}\label{eq:E_p1}
\mathbf{E}_j^{(i)} = \mathbf{A}^{(i)} + \mu_j^{(i)}\,\tilde{\mathbf{h}}_j^{(i)}(\tilde{\mathbf{h}}_j^{(i)})^\mathrm{H},
\end{align}
which removes the target-$j$ term from $\mathbf{A}^{(i)}$, so that $\mathbf{A}^{(i)} = \mathbf{E}_j^{(i)} - \mu_j^{(i)}\tilde{\mathbf{h}}_j^{(i)}(\tilde{\mathbf{h}}_j^{(i)})^\mathrm{H}$.  If $|G_j^{(i)}|^2 \ge P_\mathrm{req}$ already holds at $\mu_j^{(i)} = 0$, then $\mu_j^{(i),\star} = 0$; otherwise the floor holds with equality.  By the Sherman--Morrison formula~\cite{Bartlett51}, the inverse of $\mathbf{A}^{(i)} = \mathbf{E}_j^{(i)} - \mu_j^{(i)}\tilde{\mathbf{h}}_j^{(i)}(\tilde{\mathbf{h}}_j^{(i)})^\mathrm{H}$ is written as
\begin{align}\label{eq:sm_inv_p1}
(\mathbf{A}^{(i)})^{-1} = (\mathbf{E}_j^{(i)})^{-1} + \frac{\mu_j^{(i)}\,(\mathbf{E}_j^{(i)})^{-1}\tilde{\mathbf{h}}_j^{(i)}(\tilde{\mathbf{h}}_j^{(i)})^\mathrm{H}(\mathbf{E}_j^{(i)})^{-1}}{1-\mu_j^{(i)}\,(\tilde{\mathbf{h}}_j^{(i)})^\mathrm{H}(\mathbf{E}_j^{(i)})^{-1}\tilde{\mathbf{h}}_j^{(i)}}.
\end{align}
Multiplying $\mathbf{w}_k^{(i)} = (\mathbf{A}^{(i)})^{-1}\mathbf{b}_k^{(i)}$ by $(\tilde{\mathbf{h}}_j^{(i)})^\mathrm{H}$ and writing $\zeta_{j,k}^{(i)} = (\tilde{\mathbf{h}}_j^{(i)})^\mathrm{H}(\mathbf{E}_j^{(i)})^{-1}\mathbf{b}_k^{(i)}$ and $\omega_j^{(i)} = (\tilde{\mathbf{h}}_j^{(i)})^\mathrm{H}(\mathbf{E}_j^{(i)})^{-1}\tilde{\mathbf{h}}_j^{(i)}$, the two terms of \eqref{eq:sm_inv_p1} collapse to
\begin{align}\label{eq:sm_p1}
(\tilde{\mathbf{h}}_j^{(i)})^\mathrm{H}\mathbf{w}_k^{(i)} = \zeta_{j,k}^{(i)} + \frac{\mu_j^{(i)}\omega_j^{(i)}\,\zeta_{j,k}^{(i)}}{1-\mu_j^{(i)}\omega_j^{(i)}} = \frac{\zeta_{j,k}^{(i)}}{1-\mu_j^{(i)}\omega_j^{(i)}}.
\end{align}
The beampattern gain therefore becomes $|G_j^{(i)}|^2 = \big(\sum_{k\in\mathcal{K}}|\zeta_{j,k}^{(i)}|^2\big)/(1-\mu_j^{(i)}\omega_j^{(i)})^2$, and setting it equal to $P_\mathrm{req}$ with the root that preserves $\mathbf{A}^{(i)} \succ \mathbf{0}$ (that is, $\mu_j^{(i)}\omega_j^{(i)} < 1$) gives the closed-form update
\begin{align}\label{eq:mu_update}
\mu_j^{(i)} = \frac{1}{\omega_j^{(i)}}\Bigg(1 - \sqrt{\frac{\sum_{k\in\mathcal{K}}|\zeta_{j,k}^{(i)}|^2}{P_\mathrm{req}}}\,\Bigg)^{\!\!+},
\end{align}
where $(x)^+ = \max(x,0)$, so the multiplier update becomes positive only when the beampattern floor is violated at $\mu_j^{(i)} = 0$. After each sweep over $\{\mu_j^{(i)}\}$, $\lambda$ is recomputed from \eqref{eq:lambda_search}. By preventing target directions from being placed in beamformer nulls, cycling Steps~1--4 yields a nondecreasing objective in the FP surrogate by construction~\cite{Shen18_FP}; hence, the inner iterates converge to a stationary point of \eqref{eq:p1_problem} for fixed PA positions.

% \paragraph{Algorithm 1 (FP beamformer solver).}  Iterate: (i)~$\eta_k^{(i)} \!\gets\! \gamma_k^{(i)}$; (ii)~$\xi_k^{(i)}$ via \eqref{eq:xi_opt_p1}; (iii) bisect the shared $\lambda$ over all subcarriers and update $\{\mu_j^{(i)}\}$ by cyclic Sherman--Morrison; (iv)~$\mathbf{w}_k^{(i)}$ via \eqref{eq:wk_closed_form_p1}.  The FP block is monotone in $R_\mathrm{sum}$ \cite{Shen18_FP}; convergence in $T_\mathrm{FP} \!=\! 3$ outer iterations on average, with the per-subcarrier closed forms computed in parallel across $N_\mathrm{c}$ for each $\lambda$, at subcarrier cost $\mathcal{O}(N^3(K\!+\!J))$.

\subsection{Pinching Antenna Position Optimization}
\label{subsubsec:x_subproblem}
With $\mathbf{W}$ fixed, the PA-placement subproblem is to maximize $R_\mathrm{sum}(\mathbf{X};\mathbf{W})$ subject to the feasible set $\mathcal{X}$.  The objective is non-convex and the variables $\{x_m^{(n)}\}$ are tightly coupled through $\tilde{\mathbf{h}}$, but the feasible set has a particularly favorable per-coordinate structure: for fixed $\{x_{m'}^{(n)}\}_{m'\ne m}$, the single PA $x_m^{(n)}$ feasibly lies in the interval
\begin{align}\label{eq:x_interval}
\mathcal{F}_m^{(n)} = [\,x_{m-1}^{(n)} + \delta\,,\, x_{m+1}^{(n)} - \delta\,],
\end{align}
with the convention $x_0^{(n)} = 0$ and $x_{M+1}^{(n)} = L$. This per-coordinate feasible interval enables a one-dimensional coordinate-descent update. Specifically, for each PA $(n,m)$, the feasible interval $\mathcal{F}_m^{(n)}$ is discretized into $Q$ uniformly spaced candidate positions. The beamforming subproblem is then re-solved for each candidate position, and the candidate yielding the highest achievable sum rate is selected. Algorithm~\ref{alg:phase1_x} summarizes the resulting update, which requires $NMQ$ candidate evaluations per outer iteration.

\begin{algorithm}[t]
    \caption{Phase-1 PA-placement update via element-wise coordinate descent}
    \label{alg:phase1_x}
    \begin{algorithmic}[1]
        \Require Current $\mathbf{X}, \mathbf{W}$; grid size $Q$
        \For{$n = 1$ to $N$, $m = 1$ to $M$}
            \State Construct feasible interval $\mathcal{F}_m^{(n)}$ via \eqref{eq:x_interval}
            \State Discretise $\mathcal{F}_m^{(n)}$ into $Q$ uniform candidates $\{c_1,\ldots,c_Q\}$
            \For{$q = 1$ to $Q$}
                \State $x_m^{(n)} \gets c_q$; rebuild $\tilde{\mathbf{h}}_k^{(i)}, \tilde{\mathbf{h}}_j^{(i)}$ for all $k, j$
                \State Recompute beamformer $\mathbf{W}^{(q)}$ for the candidate channel
                \State Evaluate $R_\mathrm{sum}^{(q)} = R_\mathrm{sum}(\mathbf{X}, \mathbf{W}^{(q)})$
            \EndFor
            \State $x_m^{(n)} \gets c_{q^\star}$ with $q^\star = \arg\max_q R_\mathrm{sum}^{(q)}$
        \EndFor
        \State \Return updated $\mathbf{X}, \mathbf{W}$
    \end{algorithmic}
\end{algorithm}

\subsection{Overall Alternating Optimization}
\label{subsubsec:p1_overall}
Phase~1 alternates between the beamforming and PA-placement updates until the relative change in the achievable sum rate falls below $\epsilon_\mathrm{AO}$. Since each subproblem is solved to monotonically improve $R_\mathrm{sum}$, the sequence ${R_\mathrm{sum}^{(t)}}$ is non-decreasing and upper-bounded, and therefore converges~\cite{Bertsekas99}. The beamforming update requires $T_\mathrm{FP}N_\mathrm{c}\mathcal{O}(N^3(K+J))$ operations per outer iteration, whereas the PA-placement update incurs an additional factor of $NMQ$ due to the exhaustive evaluation of candidate positions. In the considered simulations, the outer AO converges within only a few iterations.

% \begin{algorithm}[t]
%     \caption{Phase 1: Joint PA + Beamforming Optimization (proposed pipeline)}
%     \label{alg:phase1_overall}
%     \begin{algorithmic}[1]
%         \Require Uniform initial $\mathbf{X}^{(0)}$; tolerance $\epsilon_\mathrm{AO}$
%         \State $t \gets 0$
%         \Repeat
%             \State $\mathbf{W}^{(t+1)} \gets$ \textbf{Algorithm 1} (FP beamformer solver) with $\mathbf{X}^{(t)}$ fixed
%             \State $(\mathbf{X}^{(t+1)}, \mathbf{W}^{(t+1)}) \gets$ \textbf{Algorithm 2} (element-wise PA-placement update)
%             \State $t \gets t+1$
%         \Until $|R_\mathrm{sum}^{(t)} - R_\mathrm{sum}^{(t-1)}|/R_\mathrm{sum}^{(t-1)} < \epsilon_\mathrm{AO}$
%         \State \Return $\mathbf{X}^\star = \mathbf{X}^{(t)}, \mathbf{W}^\star = \mathbf{W}^{(t)}$
%     \end{algorithmic}
% \end{algorithm}

\section{Phase 2: Target Localization}
\label{sec:phase2}

\begin{figure*}[t]
\centering
\includegraphics[width=\textwidth]{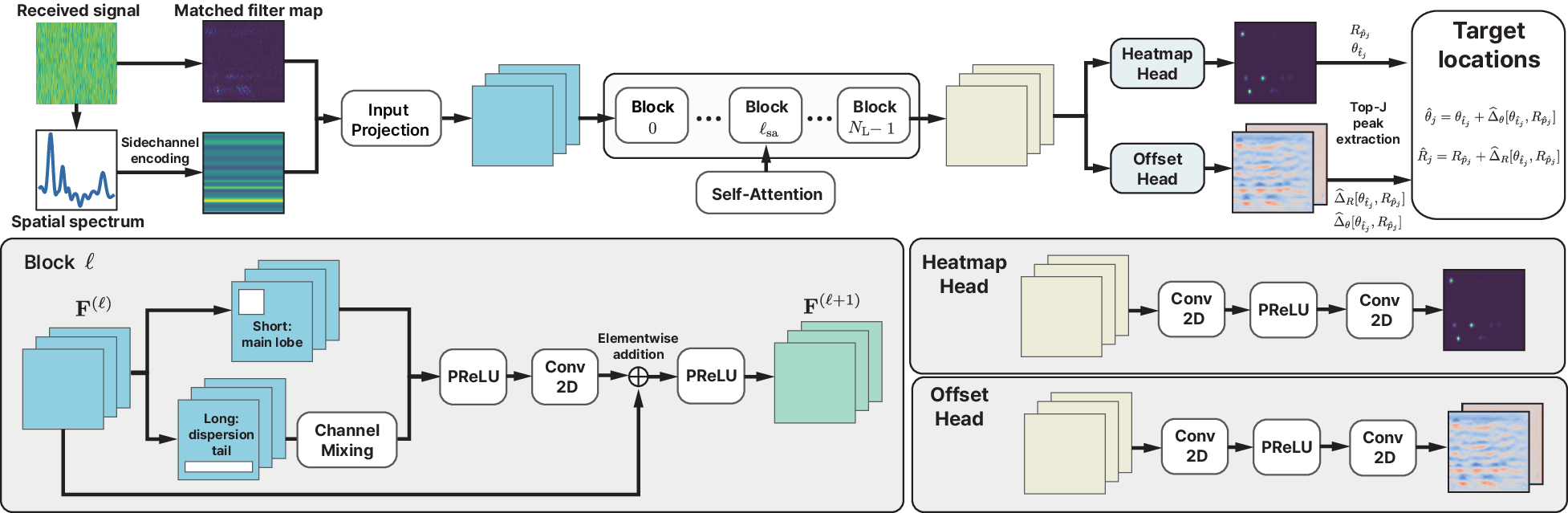}
\caption{Phase-2 DiPL-Net pipeline. The received signal is mapped to a 2D score map $\boldsymbol{\chi}$ and an auxiliary spatial-spectrum input $P_\mathrm{ss}$, which are fused, lifted by an input projection, processed by $N_\mathrm{L}$ dispersion-matched dual-kernel residual blocks with a mid-depth self-attention layer, decoded by the heatmap and offset heads, and converted into target locations via top-$J$ peak extraction.}
\label{fig:pipeline}
\end{figure*}

Phase~2 estimates the target positions $\{\hat{\mathbf{p}}_j\}_{j=1}^J$ from the received sensing signal $\mathbf{Y} \!\in\! \mathbb{C}^{N_\mathrm{R}\!\times\!N_\mathrm{c}}$ in \eqref{eq:y_matrix}. Since the detector operates after transmit beamforming, the optimized beamformer and PA placement determine the effective sensing channel in \eqref{eq:y_matrix} and \eqref{eq:Hch_def}. Phase~1 does not update the detector parameters; rather, it improves the detector input by shaping the transmit beampattern and avoiding target-direction nulls. DiPL-Net is then used as a physics-guided learning model for range-axis dispersion mitigation, as summarized in Fig.~\ref{fig:pipeline}.

\subsection{From the Sensing Observation to the Network Inputs}
\label{subsec:y_to_inputs}
\paragraph{2D matched-filter score map.}  A coarse angle grid $\{\theta_t\}_{t=1}^{D_\theta}$ over the Rx field of view and a fine range grid $\{R_p\}_{p=1}^{D_R}$ over the BS-reachable interval are fixed in advance.  For each grid angle $\theta_t$ and subcarrier $i$, correlating the observation \eqref{eq:y_matrix} with the steering vector $\mathbf{a}_\mathrm{R}(\theta_t)$ yields the angle-$\theta_t$ signal
\begin{align}\label{eq:per_beam_y}
\mathbf{y}_{\theta_t}[i] \;=\; \frac{1}{\sqrt{N_\mathrm{R}}}\,\mathbf{a}_\mathrm{R}^\mathrm{H}(\theta_t)\,[\mathbf{Y}]_{:, i},
\end{align}
which is matched against a dispersion-aware range dictionary $\boldsymbol{\Psi}_\mathrm{R}(\theta_t;\mathbf{s}) \!\in\! \mathbb{C}^{N_\mathrm{c}\times D_R}$ whose columns encode the effective response at each candidate range,
\begin{align}
\big[\boldsymbol{\Psi}_\mathrm{R}(\theta_t;\mathbf{s})\big]_{i, p} &= \sum_{k\in\mathcal{K}} h_k^\mathrm{eff}(R_p, \theta_t; f_i)\, s_k^{(i)}\, e^{-j 2\pi i\Delta f\cdot 2R_p/c}, \label{eq:p2_atom}\\
h_k^\mathrm{eff}(R, \theta; f_i) &= \big(\tilde{\mathbf{h}}(\mathbf{X}^\star, f_i, \mathbf{p}(R;\theta))\big)^\mathrm{H}\mathbf{w}_k^{\star(i)}, \label{eq:p2_h_eff}
\end{align}
with unit-norm columns.  The matched-filter correlation forms the 2D score map
\begin{align}\label{eq:p2_2d_score}
[\boldsymbol{\chi}]_{\theta_t, R_p} \;=\; \big|\big[\boldsymbol{\Psi}_\mathrm{R}(\theta_t;\mathbf{s})\big]_{:, p}^\mathrm{H}\, \mathbf{y}_{\theta_t}\big|^2, 
\end{align}
with $\mathbf{y}_{\theta_t}$ also unit-normalized so $\boldsymbol{\chi} \!\in\! [0,1]$.  In the ULA-only limit, the 2D score map establishes peaks at the true $(\theta_j, R_j)$ pairs, whereas PASS dispersion leaves each response with an extended range-axis tail whose sidelobes can mask weak targets. 

\remark The dictionary in \eqref{eq:p2_atom}--\eqref{eq:p2_h_eff} can be constructed for any given beamformer and PA layout. The Phase-1 optimized beamformer $\mathbf{W}^\star$ and PA placement $\mathbf{X}^\star$ define the effective sensing channel used in Phase~2 and improve the sensing SNR of the matched-filter score map in \eqref{eq:p2_2d_score}. This establishes the coupling between the joint beamforming/PA-placement optimization in Phase~1 and the localization stage in Phase~2.

\paragraph{Spatial-spectrum sidechannel.}  In parallel, the forward-backward-smoothed sample covariance
\begin{align}\label{eq:fb_cov}
\widehat{\mathbf{R}}_\mathrm{fb} \;=\; \tfrac{1}{2}\!\big(\mathbf{Y}\mathbf{Y}^\mathrm{H}/N_\mathrm{c} + \mathbf{J}_{N_\mathrm{R}}\,(\mathbf{Y}\mathbf{Y}^\mathrm{H}/N_\mathrm{c})^*\,\mathbf{J}_{N_\mathrm{R}}\big),
\end{align}
with $\widehat{\mathbf{R}}_\mathrm{fb} \!\in\! \mathbb{C}^{N_\mathrm{R}\times N_\mathrm{R}}$ and $\mathbf{J}_{N_\mathrm{R}}$ the $N_\mathrm{R}\!\times\!N_\mathrm{R}$ exchange matrix~\cite{Pillai89_FB}, feeds the spatial spectrum
\begin{align}\label{eq:ss_spectrum}
P_\mathrm{ss}[\theta_t] \;=\; \mathbf{a}_\mathrm{R}^\mathrm{H}(\theta_t)\,\widehat{\mathbf{R}}_\mathrm{fb}\,\mathbf{a}_\mathrm{R}(\theta_t),
\end{align}
stacked across $t$ as $P_\mathrm{ss} \in \mathbb{R}^{D_\theta}$.
$P_\mathrm{ss}$ provides an angular prior that helps separate close-angle target clusters from a single strong target response with a long range-axis dispersion tail.

\paragraph{Sidechannel encoding.} The raw spatial spectrum $P_\mathrm{ss}$ provides an angle-axis power profile whose occupied directions appear as broad peaks with angular sidelobes. Before fusion with the score map, this angle prior is refined by a lightweight three-layer angle-axis CNN $\mathrm{AE}(\cdot; \boldsymbol{\theta}_\mathrm{AE})$ with PReLU non-linearity~\cite{He15_PReLU}. Its output is given by
\begin{align}\label{eq:dipl_ae}
\mathbf{e}_\mathrm{ang} = \mathrm{AE}(P_\mathrm{ss}; \boldsymbol{\theta}_\mathrm{AE}),
\end{align}
with $\mathbf{e}_\mathrm{ang} \in \mathbb{R}^{1 \times D_\theta}$. Since $\mathbf{e}_\mathrm{ang}$ contains only angular information, it is replicated along the range dimension before being fused with the score map.

\subsection{DiPL-Net Forward Pass}
\label{subsec:dipl_arch}

\paragraph{Input fusion and projection.} The angle encoding $\mathbf{e}_\mathrm{ang}$ and the score map $\boldsymbol{\chi}$ are concatenated along the channel axis:
\begin{align}\label{eq:dipl_fuse}
\mathbf{F}_\mathrm{in} = \mathrm{concat}_c\!\big(\boldsymbol{\chi},\, \mathbf{e}_\mathrm{ang}\big)\in\mathbb{R}^{2\times D_\theta\times D_R}.
\end{align}
The fused tensor $\mathbf{F}_\mathrm{in}$ is first projected into a $C$-channel feature space through an input projection module composed of two convolutional layers with $7\!\times\!7$ and $5\!\times\!5$ kernels, each followed by a PReLU activation. The projected feature is represented by
\begin{align}\label{eq:dipl_inproj}
\mathbf{F}^{(0)}
=
\mathrm{InProj}(\mathbf{F}_\mathrm{in};\boldsymbol{\theta}_\mathrm{In})\in\mathbb{R}^{C\times D_\theta\times D_R}.
\end{align}
Since the projection kernel matches the main-lobe support of the matched-filter response, $\mathbf{F}^{(0)}$ contains local main-lobe information before the residual stack processes the dispersion.

\paragraph{Dispersion-matched residual stack.}  The dispersion physics is embedded into the backbone through an architectural inductive bias. Each residual block consists of two parallel range-axis branches: a short-kernel branch for capturing the main lobes of target response and a long-kernel branch for modeling the extended waveguide-dispersion tail. Define one pre-activation dual-kernel block as
\begin{align}\label{eq:residual_block}
\mathrm{Res}_\ell(\mathbf{F}) = {}& \mathrm{PReLU}\!\Big(\mathbf{F} + \mathbf{C}_\ell^{(2)} * \mathrm{PReLU}\big(\,\overbrace{\mathbf{C}_\ell^{\mathrm{s}} * \mathbf{F}}^{\text{short: main lobe}} \nonumber\\
& {}+ \underbrace{\mathbf{C}_\ell^{\mathrm{m}} * (\mathbf{g}_\ell \circledast_\mathrm{R} \mathbf{F})}_{\text{long: dispersion tail}}\,\big)\Big),
\end{align}
with a short kernel $\mathbf{C}_\ell^{\mathrm{s}}\!\in\!\mathbb{R}^{C\times C\times \kappa_\mathrm{s}\times \kappa_\mathrm{s}}$ sized to the range main lobe, a depthwise range-axis long kernel $\mathbf{g}_\ell\!\in\!\mathbb{R}^{C\times 1\times \kappa_\ell}$ ($\circledast_\mathrm{R}$ denotes range-axis convolution) whose length $\kappa_\ell$ spans the dispersion-tail support, a $1\!\times\!1$ channel mixing $\mathbf{C}_\ell^{\mathrm{m}}$, an output kernel $\mathbf{C}_\ell^{(2)}\!\in\!\mathbb{R}^{C\times C\times 3\times 3}$, and $\mathrm{PReLU}(\cdot)$ the activation.  The stack applies $N_\mathrm{L}$ such blocks with a self-attention layer $\mathrm{SA}(\cdot)$ inserted at depth $\ell_\mathrm{sa} \!=\! N_\mathrm{L}/2$:
\begin{align}\label{eq:dipl_stack}
\mathbf{F}^{(\ell+1)} = \begin{cases}
    \mathrm{Res}_\ell(\mathbf{F}^{(\ell)}) + \mathrm{SA}(\mathrm{Res}_\ell(\mathbf{F}^{(\ell)})), & \ell = \ell_\mathrm{sa},\\
    \mathrm{Res}_\ell(\mathbf{F}^{(\ell)}), & \text{otherwise},
\end{cases}
\end{align}
for $\ell = 0, \ldots, N_\mathrm{L}-1$, with all tensors $\mathbf{F}^{(\ell)}$ having size $\mathbb{R}^{C \times D_\theta \times D_R}$. The long branch provides a dispersion-scale range receptive field at each layer, allowing the block to suppress the dispersion tail while the short branch preserves the main lobe and refines close-angle clusters. The long-kernel length $\kappa_\ell$ is selected to cover the maximum dispersion spread observed over the adopted operating bandwidth, and the self-attention layer at $\ell_\mathrm{sa}$ provides non-local angular consistency.

\paragraph{Detection heads.}  Two head operators map the final feature $\mathbf{F}^{(N_\mathrm{L})}$ to per-pixel outputs---the heatmap logits $\widehat{\boldsymbol{\Phi}}$ and the sub-bin offsets $\widehat{\boldsymbol{\Delta}}$:
\begin{align}
\widehat{\boldsymbol{\Phi}}        &= \mathrm{HMHead}(\mathbf{F}^{(N_\mathrm{L})}; \boldsymbol{\theta}_\mathrm{H}), \label{eq:head_hm}\\
\widehat{\boldsymbol{\Delta}} &= \mathrm{OffHead}(\mathbf{F}^{(N_\mathrm{L})}; \boldsymbol{\theta}_\mathrm{O}), \label{eq:head_off}
\end{align}
with $\widehat{\boldsymbol{\Phi}} \!\in\! \mathbb{R}^{D_\theta \times D_R}$ and $\widehat{\boldsymbol{\Delta}} \!\in\! \mathbb{R}^{2 \times D_\theta \times D_R}$, and each head a two-layer $3\!\times\!3$ convolution with PReLU non-linearity in between.  Writing the full forward pass as a single operator with learnable parameters $\boldsymbol{\Theta}_\text{D} \!=\! \{\boldsymbol{\theta}_\mathrm{AE}, \boldsymbol{\theta}_\mathrm{In}, \{\mathbf{C}_\ell^{(\cdot)}\}, \boldsymbol{\theta}_\mathrm{SA}, \boldsymbol{\theta}_\mathrm{H}, \boldsymbol{\theta}_\mathrm{O}, \mathbf{g}_\ell\}$,
\begin{align}\label{eq:dipl_e2e}
(\widehat{\boldsymbol{\Phi}}, \widehat{\boldsymbol{\Delta}}) \;=\; \mathrm{DiPL}(\boldsymbol{\chi}, P_\mathrm{ss}; \boldsymbol{\Theta}_\text{D}).
\end{align}
$\widehat{\boldsymbol{\Phi}}$ is a logit map over the $D_\theta D_R$ grid pixels. After softmax normalization, it represents the probability that each grid bin contains a target peak. The offset head $\widehat{\boldsymbol{\Delta}}$ regresses the residual $(\theta, R)$ from the selected grid bin to the true target location. This dual-head structure separates discrete grid-bin detection from continuous sub-bin refinement.

\subsection{Training Procedure}
\label{subsubsec:training_2d}

\paragraph{Loss.}  The network is supervised with a softmax cross-entropy on the heatmap against a soft label $\boldsymbol{\Phi}^\star[\theta_t, R_p]$ centered at each $(\theta_j, R_j)$ and an L1 loss on the offset head at the bin nearest each target.  The total loss function is given by
\begin{align}\label{eq:dipl_loss}
\mathcal{L}_\mathrm{total}(\boldsymbol{\Theta}_\text{D}) =\; & -\sum_{t,p} \boldsymbol{\Phi}^\star[\theta_t, R_p] \log\mathrm{softmax}\!\big(\widehat{\boldsymbol{\Phi}}\big)[\theta_t, R_p] \nonumber\\
& + \mathcal{L}_\mathrm{off}(\widehat{\boldsymbol{\Delta}}, \boldsymbol{\Delta}^\star).
\end{align}
Here, the soft label is a normalized Gaussian mixture centered at the targets~\cite{Zhou19_arxiv},
\begin{align}\label{eq:dipl_label}
\boldsymbol{\Phi}^\star[\theta_t, R_p] = \frac{1}{Z}\sum_{j=1}^{J} \exp\!\bigg(\!-\frac{(\theta_t-\theta_j)^2}{2\sigma_\theta^2} - \frac{(R_p-R_j)^2}{2\sigma_R^2}\bigg),
\end{align}
with per-axis widths $(\sigma_\theta, \sigma_R)$ of one grid cell and $Z$ chosen so that $\sum_{t,p}\boldsymbol{\Phi}^\star[\theta_t, R_p] = 1$.  The offset term is a sub-bin L1 regression evaluated only at the bin $(\theta_{t_j}, R_{p_j})$ nearest each target, which is given by
\begin{align}\label{eq:dipl_offloss}
\mathcal{L}_\mathrm{off}(\widehat{\boldsymbol{\Delta}}, \boldsymbol{\Delta}^\star) = \frac{1}{J}\sum_{j=1}^{J} \big\|\widehat{\boldsymbol{\Delta}}[\theta_{t_j}, R_{p_j}] - \boldsymbol{\Delta}_j^\star\big\|_1,
\end{align}
where $\boldsymbol{\Delta}_j^\star = (\theta_j - \theta_{t_j},\, R_j - R_{p_j})$ is the true sub-bin offset of target $j$ from its nearest bin center.

% \paragraph{Stress curriculum and EMA.}  Each training applies a randomly-drawn stress level $s \!\in\! [0, s_\mathrm{max}(t)]$ that scales the magnitude of all five stressors of Sec.~\ref{subsec:sim_setup}, and the upper bound $s_\mathrm{max}(t)$ ramps from clean to full over training.  This emulates a continuous range of operating conditions in every minibatch and prevents the network from over-fitting a single noise regime.  Exponential moving-average weights are kept alongside the trained weights, which suppresses late-stage step noise; only the EMA copy is used at evaluation.

% \paragraph{Data generation.} Dataset is generated by the physics simulator: target positions uniform in the service area with a fraction in close-range clusters, RCS uniform in $[\sigma_\mathrm{min}, \sigma_\mathrm{max}]\,\mathrm{m}^2$, complex reflectivity $\alpha_j = e^{j\phi_j}\sqrt{\sigma_j}/R_j$, Rician NLoS realizations seeded by target position, transmit data symbols re-drawn per scene, and an independent noise realization per minibatch.  No measured data is required.

\subsection{Top-$J$ Peak Extraction}
\label{subsec:dipl_infer}

The number of targets $J$ is assumed known to facilitate comparison with existing localization methods. The final operator extracts $J$ predicted positions from $(\widehat{\boldsymbol{\Phi}}, \widehat{\boldsymbol{\Delta}})$ using non-maximum suppression (NMS) on the heatmap and per-pixel sub-bin offset refinement. For $j = 1, \ldots, J$, the $j$-th peak bin is selected as
\begin{align}\label{eq:infer_peak}
(\hat t_j, \hat p_j) = \arg\max_{(t,p) \in \mathcal{S}_{j-1}^c}\; \widehat{\boldsymbol{\Phi}}[\theta_t, R_p],
\end{align}
refined into continuous coordinates by the offsets:
\begin{align}\label{eq:infer_offset}
\hat\theta_j &= \theta_{\hat t_j} + \widehat{\Delta}_\theta[\theta_{\hat t_j}, R_{\hat p_j}],\\
\hat R_j &= R_{\hat p_j} + \widehat{\Delta}_R[\theta_{\hat t_j}, R_{\hat p_j}],
\end{align}
where $\mathcal{S}_{j-1}^c$ denotes the complement of the union of small NMS neighborhoods around the first $j\!-\!1$ peaks already extracted, so the $J$ peaks are mutually non-overlapping.

\section{Simulation Results}
\label{sec:simulation}

\begin{figure}[t]
\centering
\includegraphics[width=\columnwidth]{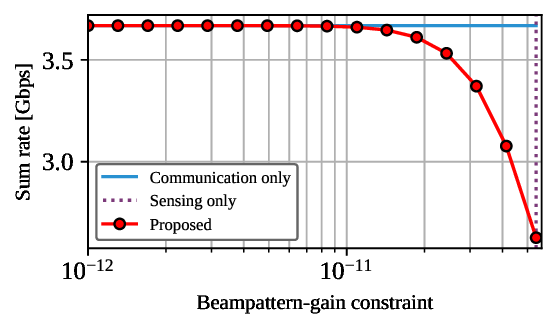}
\caption{Sum rate versus beampattern-gain constraint at the Phase-1 optimized PA placement.}
\label{fig:va_beamforming}
\end{figure}

\begin{figure}[t]
\centering
\includegraphics[width=\columnwidth]{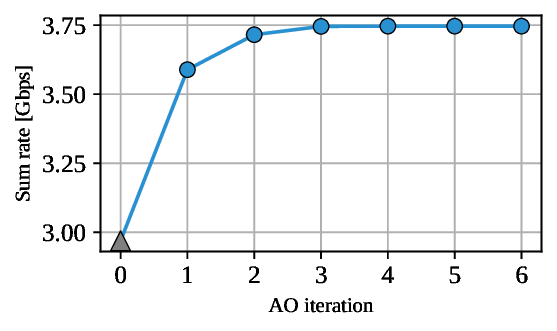}
\caption{Phase-1 alternating-optimization convergence: downlink sum rate after each joint beamformer and PA-position update.}
\label{fig:phase1_convergence}
\end{figure}

\begin{figure}[t]
\centering
\includegraphics[width=\columnwidth]{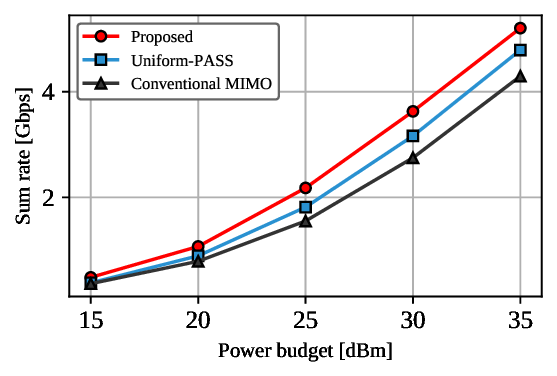}
\caption{Sum rate versus BS power budget.}
\label{fig:va_sweep_P}
\end{figure}

\begin{figure}[t]
\centering
\includegraphics[width=\columnwidth]{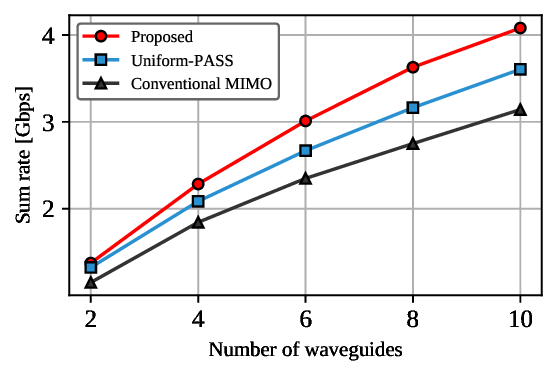}
\caption{Sum rate versus the number of waveguides.}
\label{fig:va_sweep_N}
\end{figure}

\subsection{Setup}
\label{subsec:sim_setup}
\paragraph{System parameters.} The simulation setup follows~\cite{Li26_TWC_PASS_ISAC} and extends it to wideband OFDM. Unless otherwise stated, we set the carrier frequency, bandwidth, subcarrier count, and waveguide cutoff to $f_\text{c} = 28$~GHz, $B = 400$~MHz, $N_\text{c} = 512$, and $f_\text{cut} = 26$~GHz, respectively. The Tx PASS deploys $N = 8$ waveguides uniformly over $y \!\in\! [-7.5, 7.5]$~m at ceiling height $d_h = 3$~m; each waveguide has $M = 4$ PAs and length $L = 15$~m. A co-located $N_\mathrm{R} = 16$-element half-wavelength ULA is used for reception. Users and targets are drawn in the service area $x \!\in\! [5, 20]$~m and $y \!\in\! [-7.5, 7.5]$~m, with $K = 3$ users and $J=4$ targets. We use transmit power $P_\text{tot} = 30$~dBm, receiver noise variance $\sigma_\text{s}^2 = -80$~dBm, and Rician factor $K_\mathrm{R} = 10$~dB for all effective channels. The target RCS is uniformly drawn from $\sigma_j \!\in\! [0.1, 10]\,\mathrm{m}^2$.

\paragraph{Model parameters.} DiPL-Net uses $N_\mathrm{L}\!=\!12$ dispersion-matched dual-kernel residual blocks with channel width $C\!=\!96$, and a self-attention layer is inserted at the middle of the backbone. Each block pairs a short range kernel $\kappa_\mathrm{s}\!=\!3$ with a depthwise long range kernel $\kappa_\ell\!=\!31$, which covers the observed maximum dispersion spread of approximately 30 range bins for the adopted $400$~MHz bandwidth and $26$~GHz waveguide cutoff.

\subsection{Analysis on Phase-1 ISAC Beamforming and Placement}
\label{subsec:sim_va}
Phase~1 jointly optimizes the transmit beamformers and PA positions to maximize the downlink sum rate subject to target-direction beampattern-gain constraints. The requirement $P_\mathrm{req}$ sets the operating point of the communication--sensing trade-off, as shown in Fig.~\ref{fig:va_beamforming}. When $P_\mathrm{req}$ is below the beampattern gain of the rate-optimal beamformer, the constraint is inactive and the sum-rate loss is negligible. When $P_\mathrm{req}$ exceeds this value, the constraint becomes active and the sum rate decreases toward the sensing-oriented solution. 

The proposed alternating optimization algorithm exhibits rapid convergence despite successively updating the transmit beamformer and the PA positions at each iteration, as shown in Fig.~\ref{fig:phase1_convergence}. Although the alternating updates increase the computational cost relative to a single-step optimization, the achievable sum rate converges rapidly, with the relative variation falling below $1\%$ after only a few iterations. This demonstrates that the proposed AO algorithm is computationally practical for the considered PASS-ISAC design.

The benefits of joint beamforming and PA placement are evident across both transmit-power and waveguide-count sweeps, as reported in Figs.~\ref{fig:va_sweep_P} and~\ref{fig:va_sweep_N}. The proposed design consistently achieves the highest sum rate across the considered transmit-power budgets and number of waveguides, whereas the conventional MIMO array performs worst because its fixed antenna geometry cannot adapt to user locations. Compared with Uniform-PASS, the proposed design provides the largest gain in the low-power regime, where PA placement has a stronger impact on the effective channel and link budget. Moreover, the performance gain increases with the number of waveguides, as the additional spatial degrees of freedom offer greater flexibility for jointly optimizing the beamformer and PA positions.

\subsection{Sensing Performance Analysis}
\label{subsec:sim_vb}
In this subsection, the sensing performance of the proposed DiPL-Net is evaluated and compared with representative conventional sensing detectors under identical simulation settings. The conventional baselines include the classical 2D FFT and 2D MUSIC algorithms, three beam-wise pipelines combining ESPRIT-based angle estimation with FFT, matched filtering (ML), or LISTA-based range estimation, and the dictionary-based 2D matched filter (2D ML). These methods are compared with the proposed DiPL-Net.

\begin{figure}[t]
\centering
\includegraphics[width=\columnwidth]{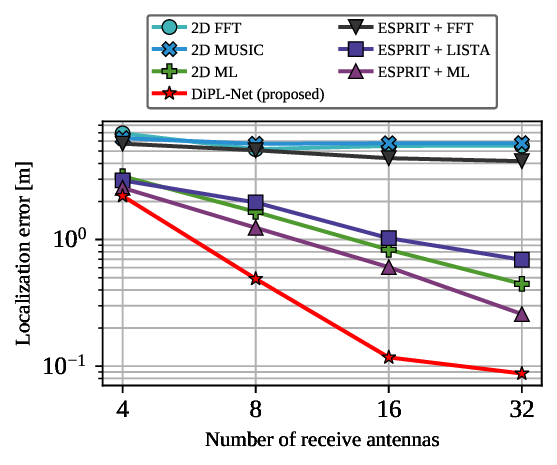}
\caption{Per-target localization error versus number of receive antennas $N_\mathrm{R}$.}
\label{fig:nr_sweep_pos}
\end{figure}

\begin{figure}[t]
\centering
\includegraphics[width=\columnwidth]{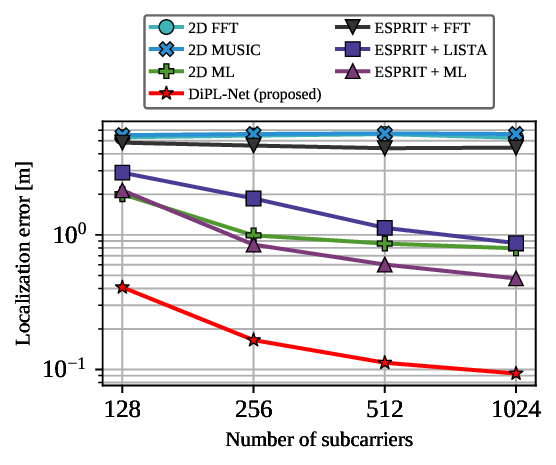}
\caption{Per-target localization error versus number of subcarriers $N_\mathrm{c}$.}
\label{fig:nc_sweep_pos}
\end{figure}

Increasing the receive aperture and the number of subcarriers reduces the localization error for all detectors, as shown in Figs.~\ref{fig:nr_sweep_pos} and~\ref{fig:nc_sweep_pos}. A larger receive aperture improves the angular resolution, whereas more subcarriers provide finer range sampling. Nevertheless, the conventional detectors remain subject to an error floor because they do not explicitly account for the waveguide-dispersion physics. In contrast, DiPL-Net consistently achieves lower localization error by embedding the underlying dispersion model into both the input representation and the network architecture, enabling effective suppression of the dispersion-induced tail while preserving the true target peak. Consequently, DiPL-Net exhibits substantially greater robustness to increasing aperture and bandwidth than the conventional baselines. 
% The performance gain eventually saturates when the localization accuracy becomes fundamentally limited by the fixed signal bandwidth.

\begin{figure}[t]
\centering
\includegraphics[width=\columnwidth]{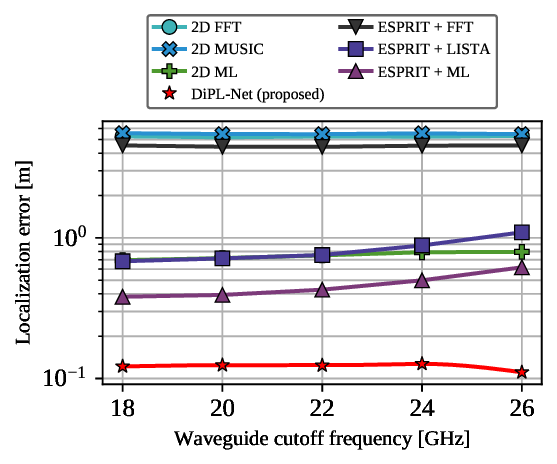}
\caption{Per-target localization error versus waveguide cutoff frequency.}
\label{fig:fcut_sensing}
\end{figure}

\begin{figure}[t]
\centering
\includegraphics[width=\columnwidth]{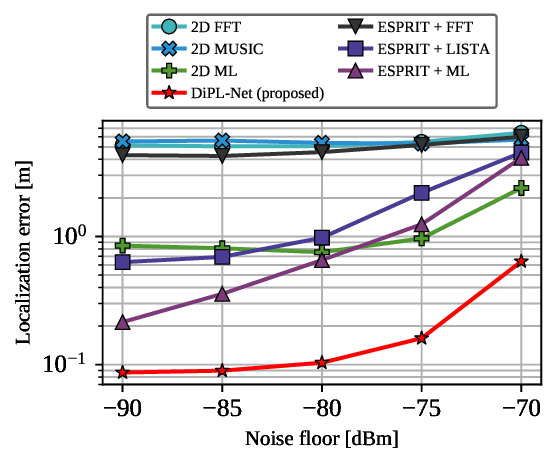}
\caption{Per-target localization error versus receiver noise floor.}
\label{fig:snr}
\end{figure}

\begin{figure}[t]
\centering
\includegraphics[width=\columnwidth]{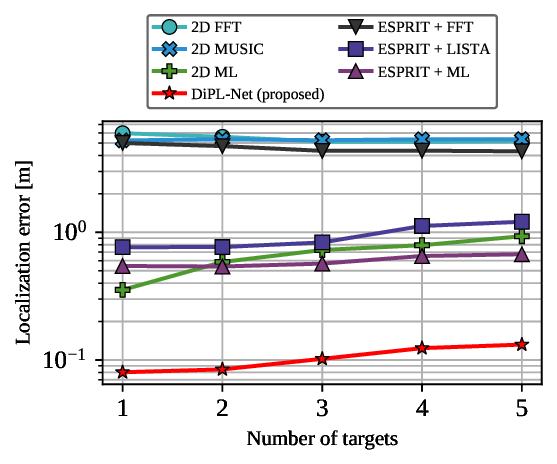}
\caption{Per-target localization error versus number of targets $J$.}
\label{fig:jsweep}
\end{figure}

\begin{figure}[t]
\centering
\includegraphics[width=\columnwidth]{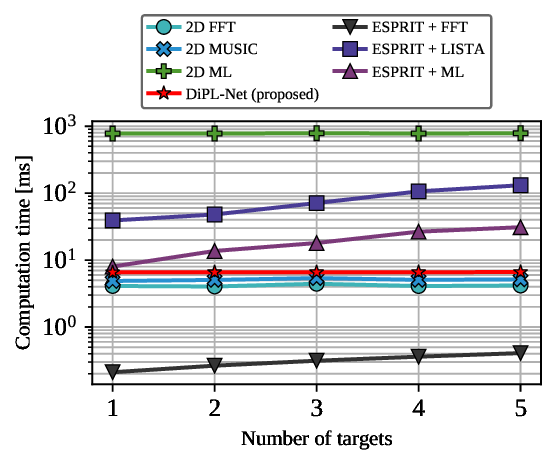}
\caption{Computation time versus number of targets for the sensing detectors.  The learned detector runs on a GPU as a single fused tensor pipeline; the classical baselines use their standard CPU implementations.}
\label{fig:timing}
\end{figure}

\begin{figure*}[t]
\centering
\includegraphics[width=\textwidth]{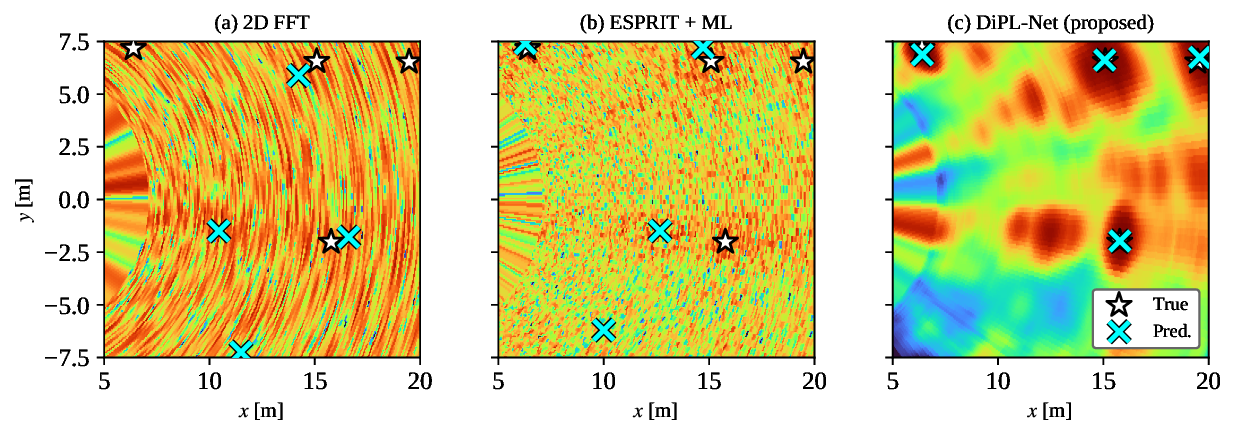}
\caption{Detection heatmaps for a multi-target scenario. (a) 2D FFT.  (b) ESPRIT + ML.  (c) proposed DiPL-Net.}
\label{fig:target_map}
\end{figure*}

The robustness evaluations demonstrate that DiPL-Net is consistently less sensitive to waveguide dispersion, receiver noise, and multi-target interference than the conventional detectors. As shown in Fig.~\ref{fig:fcut_sensing}, increasing the waveguide cutoff frequency moves the operating band closer to the cutoff edge and strengthens frequency-dependent dispersion. Since the conventional detectors do not explicitly account for this physical effect, their localization accuracy deteriorates rapidly, whereas DiPL-Net remains considerably more robust by incorporating the dispersion model into both the input representation and the network architecture. Fig.~\ref{fig:snr} further shows that DiPL-Net is more resilient to receiver noise, exhibiting a much smaller performance degradation as weak-target returns approach the noise floor. Finally, Fig.~\ref{fig:jsweep} demonstrates that increasing the number of targets intensifies overlap among dispersion-induced tails, making target separation increasingly challenging for all methods. Nevertheless, DiPL-Net consistently achieves the lowest localization error by effectively suppressing the dispersion-induced tails while preserving the true target responses.

The computational comparison in Fig.~\ref{fig:timing} shows that the proposed DiPL-Net achieves substantially better localization performance while maintaining a runtime comparable to those of the conventional sensing detectors. Since DiPL-Net predicts a heatmap over the entire angle--range grid in a single inference, its runtime remains nearly constant as the number of targets increases. By contrast, the conventional detectors typically process targets individually through repeated angle/range estimation, resulting in increasing computational time as the number of targets grows.

\begin{figure}[t]
\centering
\includegraphics[width=\columnwidth]{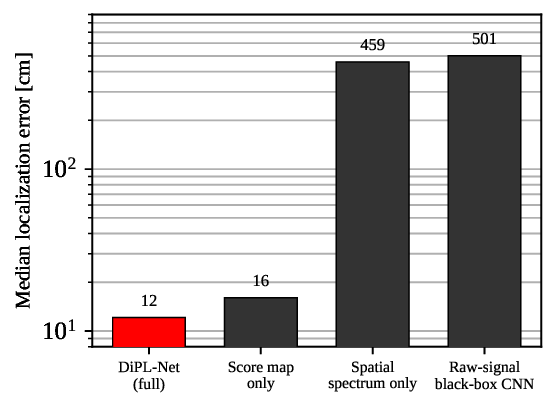}
\caption{Localization error for the same backbone trained on the full score-map-plus-spatial-spectrum input, the score map alone, the spatial-spectrum prior alone, and the raw received signal.}
\label{fig:ablation_input}
\end{figure}

Representative multi-target detection maps in Fig.~\ref{fig:target_map} illustrate the impact of waveguide dispersion on target localization. The dispersion-blind 2D FFT fails to resolve most targets because the responses are severely smeared by the dispersion-induced tails. Although ESPRIT+ML provides reasonably accurate angle estimates, it cannot separate targets sharing similar angles, and its range estimates remain degraded by waveguide dispersion, resulting in blurred responses and missed detections. By explicitly incorporating the underlying dispersion physics, DiPL-Net suppresses the dispersion-induced tails, restores compact target peaks, and successfully localizes all targets in this example.

\subsection{Ablation Study}
\label{subsec:sim_vc}

\begin{table}[t]
\caption{Component ablation of the proposed pipeline.}
\label{tab:factorial_ablation}
\centering
\resizebox{\columnwidth}{!}{%
\begin{tabular}{cccrr}
\toprule
\textbf{Beamforming} & \textbf{Position Opt.} & \textbf{Sensing Detector} & $R_\mathrm{sum}$~[Gbps] & \textbf{Pos.\ error}~[cm] \\
\midrule
$\times$     & $\times$     & $\times$     & $2.27$ & $137$ \\
$\times$     & $\times$     & $\checkmark$ & $2.27$ & $16$ \\
$\times$     & $\checkmark$ & $\times$     & $2.71$ & $117$ \\
$\times$     & $\checkmark$ & $\checkmark$ & $2.71$ & $15$ \\
\midrule
$\checkmark$ & $\times$     & $\times$     & $3.07$ & $159$ \\
$\checkmark$ & $\times$     & $\checkmark$ & $3.07$ & $17$ \\
$\checkmark$ & $\checkmark$ & $\times$     & $\mathbf{3.70}$ & $100$ \\
\rowcolor{black!8}
$\checkmark$ & $\checkmark$ & $\checkmark$ & $\mathbf{3.70}$ & $\mathbf{13}$ \\
\bottomrule
\end{tabular}}
\end{table}

The input ablation in Fig.~\ref{fig:ablation_input} evaluates the contribution of each input component to the proposed DiPL-Net. Training the same backbone directly on the raw received signal results in poor localization performance, indicating that the network alone cannot effectively learn the dispersion characteristics. Replacing the raw signal with the proposed dispersion-aware score map yields the largest performance improvement, demonstrating the importance of embedding the underlying physics into the input representation. Incorporating the auxiliary spatial-spectrum prior further improves localization accuracy by providing complementary angular information, whereas the angular prior alone is insufficient for accurate target localization because it contains no range information.

To evaluate the contribution of each module in the proposed ISAC framework, Table~\ref{tab:factorial_ablation} reports a full-factorial ablation study in which the beamforming optimization, PA-position optimization, and DiPL-Net are independently enabled or disabled. When a component is disabled, the corresponding baseline adopts maximum-ratio transmission (MRT), a uniform PA layout, or the classical 2D matched filter, respectively.

The results demonstrate that the three components play complementary roles. The optimized beamformer and PA placement improve both communication and sensing performance by jointly optimizing the propagation conditions and the effective sensing channel, resulting in higher communication sum rate and improved localization accuracy. In contrast, DiPL-Net affects only the sensing stage and therefore has negligible impact on the communication performance, while providing the dominant improvement in localization accuracy by compensating for the waveguide-dispersion effect. Consequently, enabling all three components simultaneously achieves the highest communication sum rate and the lowest localization error among all evaluated configurations.

\section{Conclusion}
\label{sec:conclusion}

This paper investigated target localization in wideband OFDM pinching-antenna systems, where waveguide dispersion severely distorts the range response and degrades localization accuracy. To address this challenge, we proposed DiPL-Net, which embeds the underlying dispersion physics into both the input representation and the network architecture to effectively suppress the dispersion-induced tails while preserving the true target responses. We further integrated DiPL-Net into a two-stage PASS-ISAC framework, where optimized beamforming and PA placement shape the effective sensing channel for subsequent localization. Simulation results demonstrated that the proposed framework significantly improves localization accuracy over conventional sensing methods while preserving the communication performance achieved by the Phase-1 optimization.

Future work includes extending the proposed framework to robust localization under imperfect target-angle information and target-location uncertainty, estimating an unknown number of targets, enabling real-time PA repositioning for dynamic environments, and incorporating higher-fidelity electromagnetic waveguide models that account for frequency-dependent leakage and radiation.

\bibliographystyle{IEEEtran}
\bibliography{bibtex}

\end{document}